\documentclass[conference]{IEEEtran}
\ifCLASSINFOpdf
  \usepackage[pdftex]{graphicx}
\else
  \usepackage[dvips]{graphicx}
\fi
\usepackage{amsmath}
\ifCLASSOPTIONcompsoc
  \usepackage[caption=false,font=normalsize,labelfont=sf,textfont=sf]{subfig}
\else
  \usepackage[caption=false,font=footnotesize]{subfig}
\fi

\usepackage{stfloats}
\usepackage{url}
\usepackage{color}
\usepackage{gensymb}
\usepackage{comment}
\usepackage{multirow}

\begin{document}
%
\title{PointNet on FPGA for Real-Time LiDAR Point Cloud Processing}

\author{
\IEEEauthorblockN{Lin Bai, Yecheng Lyu, Xin Xu and Xinming Huang}
\IEEEauthorblockA{Worcester Polytechnic Institute\\
Worcester, MA 01609, USA\\
\{lbai2, ylyu, xxu10, xhuang\}@wpi.edu}
}


\maketitle

\begin{abstract}
LiDAR sensors have been widely used in many autonomous vehicle modalities, such as perception, mapping,  and localization. This paper presents an FPGA-based deep learning platform for real-time point cloud processing targeted on autonomous vehicles. The software driver for the Velodyne LiDAR sensor is modified and moved into the on-chip processor system, while the programmable logic is designed as a customized hardware accelerator. As the state-of-art deep learning algorithm for point cloud processing, PointNet is successfully implemented on the proposed FPGA platform. Targeted on a Xilinx Zynq UltraScale+ MPSoC ZCU104 development board, the FPGA implementations of PointNet achieve the computing performance of 182.1 GOPS and 280.0 GOPS for classification and segmentation respectively. The proposed design can support an input up to 4096 points per frame. The processing time is 19.8 ms for classification and 34.6 ms for segmentation, which meets the real-time requirement for most of the existing LiDAR sensors.
\end{abstract}


%
\IEEEpeerreviewmaketitle

\section{Introduction}
Nowadays, LiDAR plays an important role in autonomous vehicle systems, due to its many advantages such as 3D information capturing capability, no environment light requirement, and etc. One or more LiDAR sensors are often installed on an autonomous vehicle for the modalities of perception\cite{zhou2018voxelnet}, mapping\cite{droeschel2018efficient}, and localization\cite{yin20193d}. One major challenge for a LiDAR system is real-time point cloud processing. 


In general, point cloud neural networks can be divided into three subcategories: pixel-based approaches, voxel-based approaches and 3D point-based approaches. Pixel-based methods project the 3D point cloud into 2D, either Bird Eye View (BEV) \cite{ku2018joint} or front view\cite{chen2017multi}. Subsequently, deep neural networks for 2D images can be applied directly. The voxel-based methods partition the 3D space into voxel grid and utilize neural networks to extract features from this grid. Both aforementioned methods may lead to information loss. The 3D point-based method, however, directly takes the raw point cloud as input in the form of $(X,Y,Z,I)$. It does not need the complicated statistics operations when comparing to the voxel-based method. Meanwhile, it avoids too much information loss when comparing to the pixel-based method. The 3D point-based methods, such as PointNet\cite{qi2017pointnet}, can produce much higher accuracy and therefore have attracted lots of research attentions.

In the paper, we propose an FPGA platform for PointNet implementations. Because of the heterogeneous architecture, Xilinx Zynq SoC chip can run the software driver for LiDAR interface on its Processing System (PS) side, and put the customized hardware accelerator on the Programmable Logic (PL) side. Data transfer between PS and PL is via Direct Memory Access (DMA).

\section{Related Work}
Several previous works \cite{shen2018towards}\cite{wu2010high} were focused on accelerating matrix multiplication using one-dimension systolic array on FPGA, which achieved efficient resource usage and low bandwidth. Newly proposed architectures \cite{zhang2015optimizing}\cite{qiu2016going} for neural networks take advantage of Single Instruction Multiple Data (SIMD) structure for matrix multiplication.
Continental AG released Assisted \& Automated Driving Control Unit (ADCU)\cite{ADCU2018}, on which a Zynq UltraScale+ MPSoC chip was loaded. It supports LiDAR processing but no technical details were revealed. NVIDIA proposed DRIVE AGX self-driving computer platforms built on Xavier SoC chip, which is capable to process point cloud data received from a LiDAR. In \cite{lyu2018real} and \cite{lyu2018chipnet}, the LiDAR was connected to a PC via Ethernet, and after pre-processing on PC, feature maps were fed to a neural network accelerator in an FPGA.


The contributions of this paper are summarized as follows:
\begin{enumerate}
    \item \textcolor{black}{To our knowledge, this is one of the first end-to-end FPGA-based platforms for point cloud deep learning}, via Ethernet. A LiDAR is connected to the PS side directly. After pre-processing by the LiDAR driver, point cloud is stored in DDR memory that is accessible to the hardware accelerator on PL side.
    \item More specifically, PointNet has been implemented on this platform as an example of point cloud deep learning algorithm. As the state-of-art deep neural network for point cloud processing, PointNet is the backbone of many latest works on 3D classifications and segmentation. Based on this, one can easily extend our implementation of PointNet accelerator to other neural networks.
    \item A scalable SIMD matrix multiplication architecture is proposed, which is capable of processing matrix in arbitrary size. This accelerator is able to process point cloud with arbitrary number of points and generate output in row order or column order. For an input of 4096 points per frame, the accelerator achieves the speed of 50.5 and 28.9 frames per second for classification and segmentation, respectively. Considering most of the LiDAR scans are at 10Hz, this accelerator fulfills the real time processing requirement.
\end{enumerate}

The rest of the paper is organized as follows: The function of LiDAR driver is described in Section ~\ref{sec:driver}. After that, the structure of PointNet is introduced in Section ~\ref{sec:pointnet}. Section ~\ref{sec:optimize}-\ref{sec:hw_arch} states the hardware optimization techniques and architecture. The evaluation results and analysis are given in Section ~\ref{sec:result}. In the end, we conclude the paper in Section ~\ref{sec:conclude}.

\section{Point Cloud Pre-processing}\label{sec:driver}
Before fed into PointNet, the raw data from LiDAR need pre-processing. We modified the driver for the embedded ARM processor on our FPGA platform. Pre-processing includes the following operations:
\begin{enumerate}
    \item Coordinate Transformation: LiDAR scans the physical world in spherical coordinate, while the PointNet requires input data in Cartesian coordinate.
    \item Time Offset: The distance to an object is measured by the time difference between emitting and receiving optics after hitting the object. However, during this small round trip time, LiDAR rotates an angle. This leads to a time offset. LiDAR driver should compensate this time difference.
\end{enumerate}

\section{PointNet}\label{sec:pointnet}

\begin{figure*}
    \centering
    \includegraphics[width=0.8\textwidth]{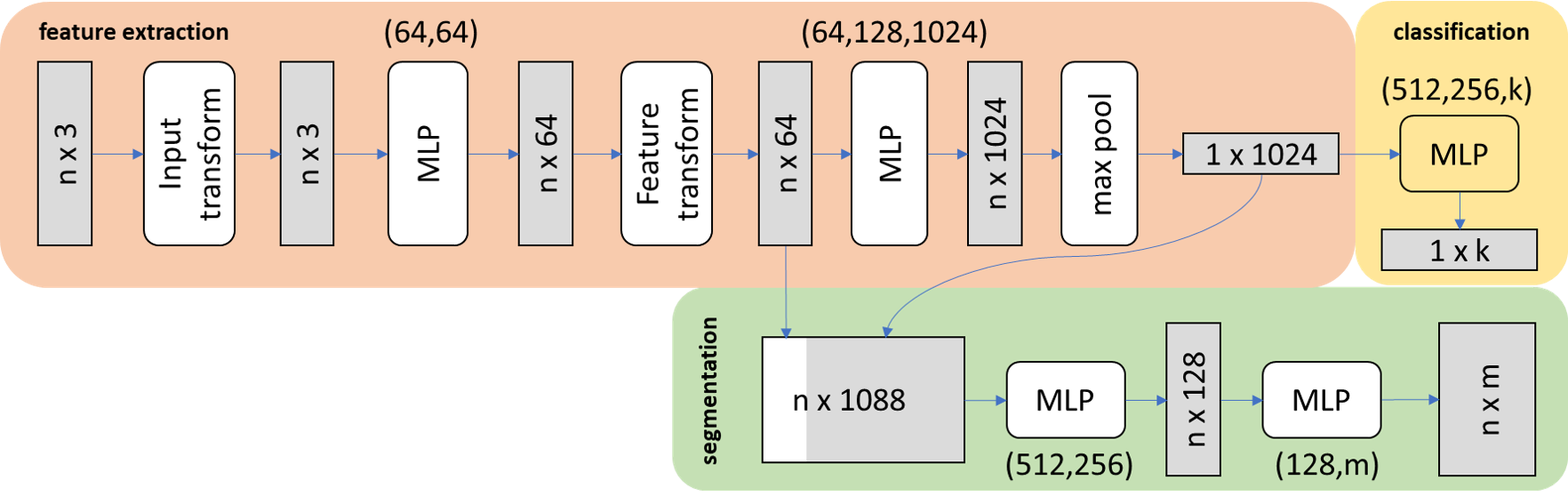}
    \caption{PointNet for point cloud classification and segmentation}
    \label{fig:point_arch}
\end{figure*}

\begin{figure}[htbp]
    \centering
    \includegraphics[width=0.8\columnwidth]{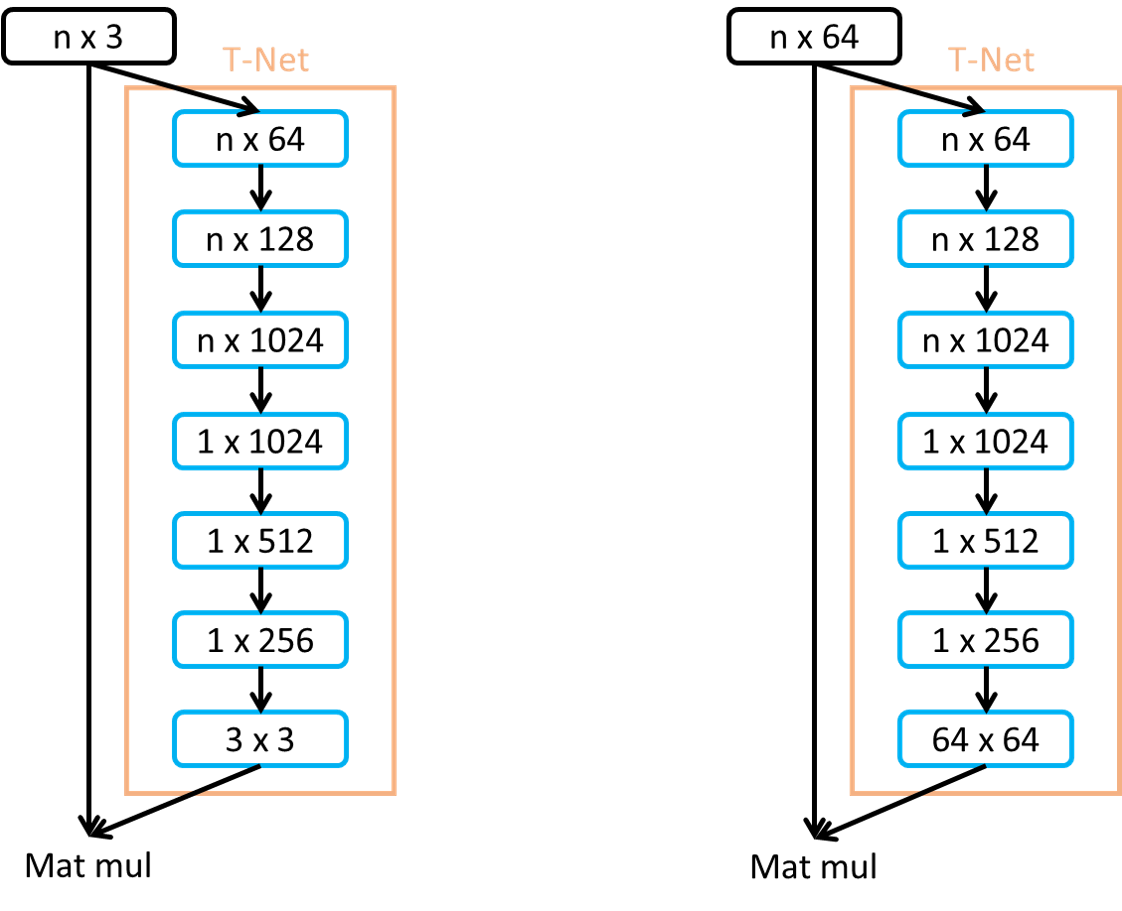}
    \caption{Transform structures in PointNet, where $n\times 3$ is for input transform and $n\times 64$ for feature transform}
    \label{fig:tnet}
\end{figure}

PointNet \cite{qi2017pointnet} is a state-of-art deep neural network algorithm that was developed for point cloud classification and segmentation. Unlike ordinary neural networks who adopt tensors as input, PointNet's input is a $n\times 3$ matrix where $n$ is the number of points and $3$ represents the position $(X,Y,Z)$ in Cartesian coordinate of one point. PointNet architecture is shown in Fig.~\ref{fig:point_arch}, where shared Multi-Layer Perceptron (MLP) is $1\times 1$ convolution mathematically. So PointNet can be realized by fully connected layers with branches. The transformation structures are illustrated in Fig.~\ref{fig:tnet} with input matrix $n\times M$, where $M=3$ in case of input transform and $M=64$ in feature transform.

As one of the most well-known deep learning algorithms for point cloud processing, PointNet is widely used as the backbone of many state-of-the-art neural networks not only for classification and segmentation, but also for object detection. For instance, PointNet is used in PointFusion \cite{xu2018pointfusion} and Attentional PointNet \cite{paigwar2019attentional} to extract point-wise feature and global feature for object detection task. It also applies to STD \cite{yang2019std} and L3-Net \cite{lu2019l3}.

\section{Optimization Strategy}\label{sec:optimize}
\subsection{Loop optimization}
\begin{figure}[htbp]
    \centering
    \includegraphics[width=0.7\columnwidth]{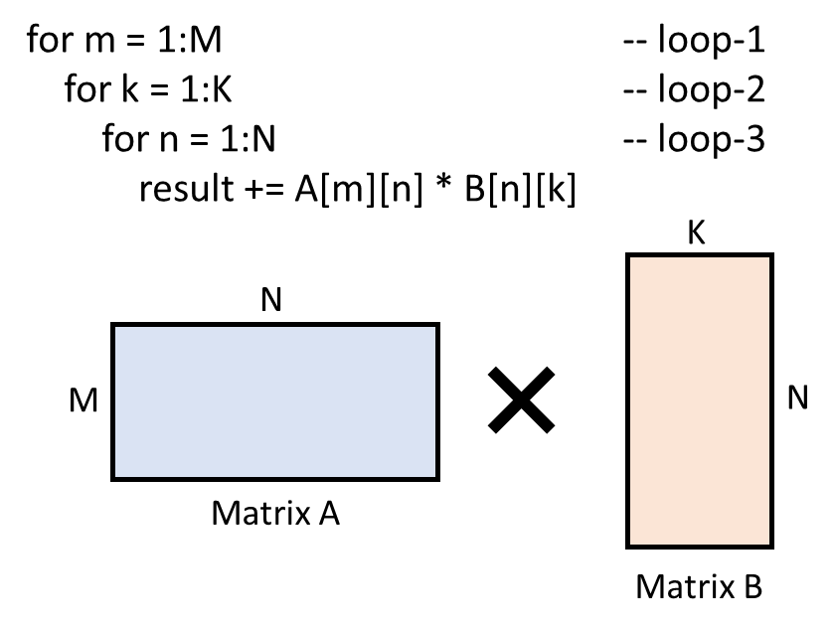}
    \caption{Loop optimization for matrix multiplication}
    \label{fig:loop_opt}
\end{figure}

From mathematical point of view, $1\times 1$ convolution is equivalent to matrix multiplication as shown in Fig.~\ref{fig:loop_opt}, which consists of 3 cascaded loops. To fully utilize the parallel processing capability of an FPGA, these loops needs to be optimized \cite{qiu2016going}\cite{ma2017optimizing} for balancing the process time and resource usage.

Loop-1: It depends on the storage of matrix. In PetaLinux or in C code, the matrix is stored in row orientation. In order to fetch data using DMA, it is not wise to unroll this loop.

Loop-2: Unrolling this loop determines how many times an accelerator has to read the input feature map. Together with loop-3, this is limited by the on-chip computation resources, i.e. DSP slices on the FPGA in our case.

Loop-3: Unrolling this loop increases the throughput of the accelerator. However, it is restricted by the communication bandwidth of HP (high performance) interface between PS and PL in Zynq. Partial unrolling of this loop leads to partial sum so that intermediate buffer becomes necessary, which can be merged into output buffer at the cost of high power consumption owning to on-chip memory access.

As for the weight matrix obtained from training, it can be pre-loaded into block RAM, so its storing and loading are flexible.

\subsection{Quantization}
A quantization during training method described in \cite{lyu2018chipnet} is adopted in this design. Avoiding the modification of TensorFlow source code, it supplies convenient solution for quantization. In this study, we quantizied PointNet parameters into 8-bit and 16-bit respectively.

\section{System Architecture of PointNet Hardware Accelerator}\label{sec:hw_arch}
\begin{figure}[htbp]
    \centering
    \includegraphics[width=0.95\columnwidth]{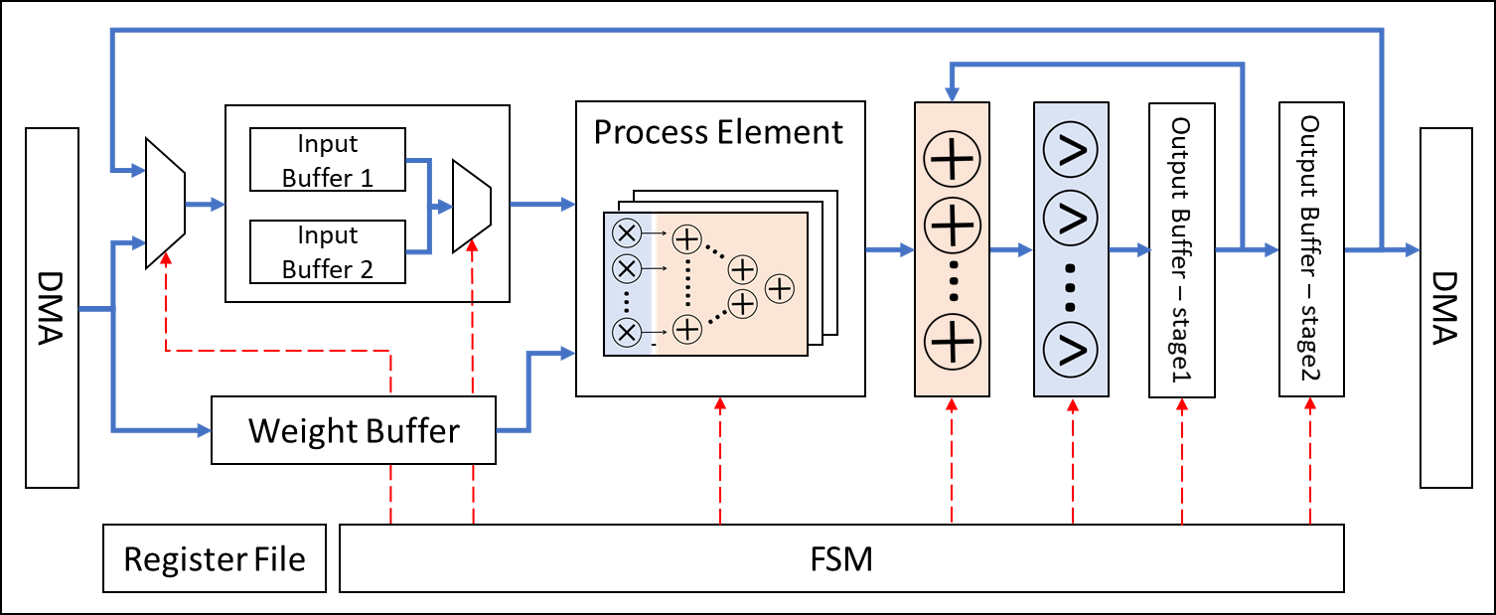}
    \caption{Hardware architecture of PointNet accelerator}
    \label{fig:point_hw_arch}
\end{figure}

Based on the description in Section~\ref{sec:pointnet}, all PointNet operations can be categorized into either matrix multiplication or max pooling. Therefore, the computing blocks involving in the PointNet accelerator (Fig.~\ref{fig:point_hw_arch}) are Process Element (PE) array for matrix multiplication, an adder array for partial sum, and a comparator array for max pooling and ReLU (Rectified Linear Unit). During inference, Batch Normalization (BN) is absorbed into the PE. Concerning to the feature map storage, double buffering technique is applied to both input buffer and weights buffer to boost the throughput. The output buffer is also designed as a two-stage buffer.

\subsection{PE Array and Buffers}
The PE consists of a multiplier array, an pipelined adder tree and an adder array. According to our loop unrolling method, loop 2 and 3 are both partially unrolled. Loop 2 partial unrolling determines the number of multipliers and the size of adder tree in each PE. Loop 3 partial unrolling factor is related to the size of weight buffer and number of PEs in the array.

\begin{figure}[htbp]
    \centering
    \includegraphics[width=0.55\columnwidth]{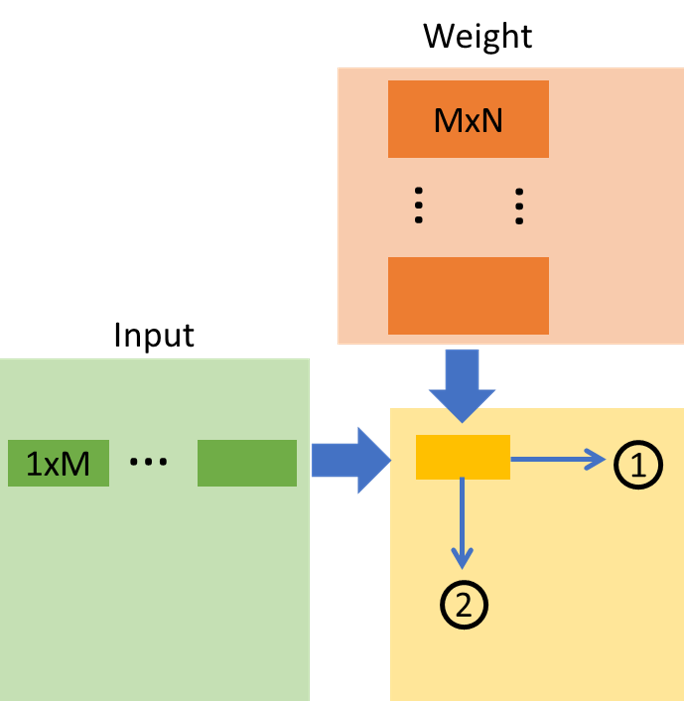}
    \caption{Matrix multiplication by block}
    \label{fig:pe_arch}
\end{figure}

As indicated as \textcircled{1} and \textcircled{2} in Fig.~\ref{fig:pe_arch}, the PE array supports both row-oriented output and column-oriented output by applying different reading patterns to the input and weight buffer.

Double buffering is designed for the input buffer and the weight buffer, so that the imbalance between processing throughput and HP port bandwidth is alleviated. Besides, a 2-stage output buffer is deployed right after the PE array. The first stage is for partial sum during matrix multiplication. It has wider bitwidth than the second stage. The second stage is for storing the final results and transferring data to DDR via DMA. This structure is designed for two reasons, one is to avoid the precision reduction introduced by matrix partitioning, the other one is to alleviate frequent reading of second stage output buffer, so partial sum accumulation and data sending to DDR can work simultaneously.

\begin{figure}[htbp]
    \centering
    \includegraphics[width=0.9\columnwidth]{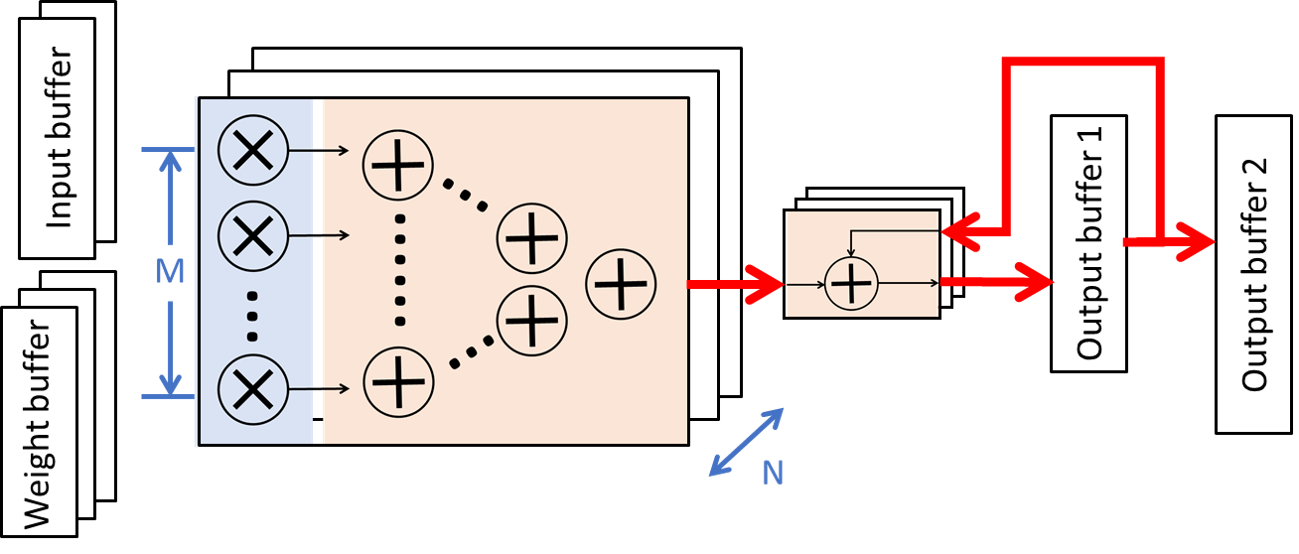}
    \caption{Structure of the processing element and the buffers}
    \label{fig:block_mmult}
\end{figure}

\subsection{Max-pooling and ReLU}
In this PointNet accelerator, max-pooling and ReLU share one comparator array. Max-pooling is to find out the largest value in each feature that each column of the matrix. In order to merge max-pooling into matrix multiplication pipeline, the output pattern is charged from row-by-row to column-by-column for the PE array. The ReLU function compares the results with 0 to filter out the negative values. Besides ReLU, other similar function like ReLU6 is also supported.

\subsection{Operation Control}
Prior to run, ARM core sends configurations to register file block, including the number of points and its destination buffer. The matrix multiplication patterns are also pre-defined and loaded into register file, which determines how the result comes out (row oriented or column oriented) and whether the result will be sent to DDR or input buffer for the next operation. According to the configurations read from register file, a FSM (finite state machine) sends control signal to each block. Double buffering enables this accelerator to accept new weights or input during processing. The FSM also handles  the assignment of two buffers, one for receiving and the other for sending. The usage of register file speeds up the processing. By pre-loading all needed parameters into the accelerator, no interrupt based configuration mechanism is necessary, which avoids the slow down due to interrupt handling in PetaLinux.

\section{Implementation Results}\label{sec:result}
The accelerator is designed using Simulink and the HDL Coder toolbox. The evaluation platform is Xilinx Zynq UltraScale+ MPSoC ZCU104 Development Kit. 
When operating in 64-bit mode, the maximum bandwidth of DDR is $102.4$Gbps at $800$MHz. The LiDAR mounted on this platform is Velodyne VLP-16.

\begin{figure}[htbp]
    \centering
    \includegraphics[width=0.9\columnwidth]{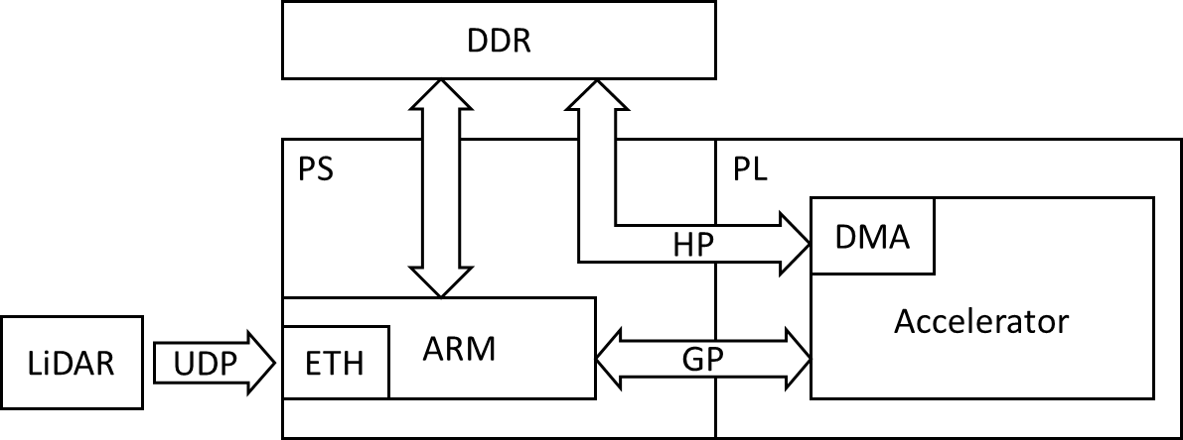}
    \caption{Overview of the LiDAR processing framework}
    \label{fig:sys_arch}
\end{figure}

Fig.~\ref{fig:sys_arch} presents the test setup of the LiDAR processing framework. PetaLinux operating system is running on the ARM core on PS side. The point cloud is received via Ethernet interface using UDP protocol. After processed by the Velodyne driver, point cloud in Cartesian coordinate ROI (region of interest) is transmitted into DDR memory. Then ARM loads setting parameters for the hardware accelerator through General Purpose (GP) port based on AXI-lite protocol. During execution, the accelerator loads or stores point cloud (or intermediate data) at DDR by DMA via High Performance (HP) port according to AXI Stream protocol.

As described in the previous sections, the matrix multiplication pattern is pre-loaded into configuration memory. Therefore, this design is able to implement the full PointNet or PointNet-vanilla, which is a simplified PointNet without transforms, for classification or segmentation task. The maximum number of points supported in this design is $4096$. Larger point cloud can be fed into this design after partitioning. For autonomous driving applications, a Velodyne VLP-16 LiDAR running at $10$Hz supplies around $360/0.2\times 16=28.8K$ points in each point cloud. Considering the normally used Region Of Interest (ROI) is a $20m\times 60m$ square in front of the vehicle (less than $1/6$), the points in ROI is less than $4096$. For high resolution LiDAR such as Velodyne HDL-64E, total number of poin
in ROI is much larger than $4096$. To be fed into this accelerator, the point cloud can be sub-sampled or partitioned.

Table~\ref{tab:bitwidth} summarizes the on-chip resources consumption when choosing matrix dimension size of $M=32$ and $N=32$ in Fig~\ref{fig:pe_arch}. In practical applications, users can choose the suitable bitwidth based on available FPGA resource and processing speed requirement.

\begin{table}[htbp]
    \centering
    \caption{FPGA resource consumption of PointNet}
    \begin{tabular}{ |c|c|c|c|c|c| } 
        \hline
        Width & LUT & FF & DSP & BRAM & URAM \\
        \hline
        \multirow{2}{*}{INT8} & 19530 & 36010 & 1026 & 114 & 48\\
                              & 8\%   & 8\%   & 60\% & 37\%& 50\%\\ 
        \hline
        \multirow{2}{*}{INT16} & 30933 & 60412 & 1026 & 123 & 96\\ 
                               & 13\%  & 13\%  & 60\% & 39\%& 100\%\\ 
        \hline
    \end{tabular}
    \label{tab:bitwidth}
\end{table}

Tab.~\ref{tab:perform} compares the throughput and processing speed in terms of different quantization bit width.
The PointNet accelerator takes 19.8 ms and 34.6 ms to classify and segment a point cloud with 4096 points respectively when using INT8 quantization. Considering most of the LiDAR scans at 10Hz, this PointNet accelerator is able to work in real time.

\begin{table}[htbp]
    \centering
    \caption{Comparison of performance}
    \begin{tabular}{|c|c|c|c|c|}
        \hline
        \multirow{2}{*}{Networks} & \multicolumn{2}{c|}{Throughput(GOPS)} &
        \multicolumn{2}{c|}{Processing time(ms)} \\
        \cline{2-5}
          & int8 & int16 & int8 & int16 \\
        \hline
        PointNet-vanilla   & \multirow{2}{*}{112.5} & \multirow{2}{*}{64.9} & \multirow{2}{*}{10.9} & \multirow{2}{*}{18.9} \\
         classification      &  &  &  & \\
        \hline
        Point-classification & 182.1 & 130.0 & 19.8 & 27.8\\
        \hline
        Point-segmentation   & 280.0 & 227.4 & 34.6 & 42.6\\
        \hline
    \end{tabular}
    \label{tab:perform}
\end{table}

\section{Conclusions}\label{sec:conclude}
In this paper, a FPGA-based LiDAR processing platform is proposed to accelerate point cloud deep learning algorithms. More specifically, a scalable PointNet hardware accelerator has been implemented on the FPGA SoC platform. For classification of an input frame with 4096 points, it only takes 19.8 ms reaching an estimated performance of about 182.1 GOPS. For segmentation task, it takes 34.6 ms per frame at the performance of about 280 GOPS. In addition, the design leaves some resource margin, so one can easily extend it for more advanced detection neural networks such as PointFusion\cite{xu2018pointfusion} and Attentional PointNet\cite{paigwar2019attentional}, or other segmentation neural networks like STD\cite{yang2019std} and L3-Net\cite{lu2019l3}.

\section*{Acknowledgment}
This work was supported by the Mathworks Inc.



%


\bibliographystyle{ieeetr}
\bibliography{lidar_platform}

\end{document}